\documentclass[%
 reprint,
 amsmath,amssymb,
 aps,
floatfix,
]{revtex4-2}

\usepackage{graphicx}
\usepackage{dcolumn}
\usepackage{bm}
\usepackage{setspace}
\usepackage[colorlinks,citecolor=blue]{hyperref}

\usepackage[dvipsnames]{xcolor}
\usepackage{url}

\newcommand{\I}{\mathrm{i}} 					

\newcommand{\Br}[1]{\ensuremath{\left( #1 \right)}} 

\newcommand{\td}{\ensuremath{\left(t\right)}} 
\newcommand{\fd}{\ensuremath{\left(\omega\right)}} 

\newcommand{\Schrod}{Schr\"{o}dinger} 

\begin{document}
\preprint{APS/123-QED}

\title{Continuous optical-to-mechanical quantum state transfer in the unresolved sideband regime}

\author{Amy Navarathna$^{1,2}$}
\author{James S. Bennett$^{1,2,3}$}
\author{Warwick P. Bowen$^{1,2}$}
\email{w.bowen@uq.edu.au}

\affiliation{
 $^1$ ARC Centre of Excellence for Engineered Quantum Systems, St Lucia, Queensland 4072, Australia\\
 $^2$ School of Mathematics and Physics, University of Queensland, St Lucia, Queensland 4072, Australia\\
 $^3$ Centre for Quantum Dynamics, Griffith University, Nathan, QLD 4222, Australia
}
\date{\today}

\begin{abstract}

Optical-to-mechanical quantum state transfer is an important capability for future quantum networks, quantum communication, and distributed quantum sensing. However, existing continuous state transfer protocols operate in the resolved sideband regime, necessitating a high-quality optical cavity and a high mechanical resonance frequency. Here, we propose a continuous protocol that operates in the unresolved sideband regime. The protocol is based on feedback cooling, can be implemented with current technology, and is able to transfer non-Gaussian quantum states with high fidelity. Our protocol significantly expands the kinds of optomechanical devices for which continuous optical-to-mechanical state transfer is possible, paving the way towards quantum technological applications and the preparation of macroscopic superpositions to test the fundamentals of quantum science.

\end{abstract}


\maketitle

The ability to transfer 
quantum states between optical communication channels and quantum computing nodes is a necessary ingredient of the emerging quantum internet \cite{kimble_quantum_2008}. Quantum state transfer also has important applications in quantum-enhanced sensing~\cite{degen_quantum_2017,lachance-quirion_entanglement-based_2020}, quantum-secure communications~\cite{riedinger_remote_2018}, and fundamental tests of macroscopic quantum mechanics~\cite{paternostro_creating_2007,pikovski_probing_2012,arndt_testing_2014,forstner_nanomechanical_2020,kotler_direct_2021,mercier_de_lepinay_quantum_2021}. A leading approach is to mediate the transfer using an optomechanical resonator~\cite{barzanjeh_reversible_2012, palomaki_coherent_2013,andrews_quantum-enabled_2015,lambert_coherent_2020,arnold_converting_2020,mirhosseini_superconducting_2020}. This is attractive because mechanical resonators interact via radiation pressure with electromagnetic fields of all frequencies~\cite{bowen_quantum_2015} and can also be functionalized to interact with most quantum computing nodes, such as spins~\cite{rabl_strong_2009,ovartchaiyapong_dynamic_2014,carter_spin-mechanical_2018}, superconducting devices~\cite{schneider_coupling_2012,arrangoiz-arriola_coupling_2018,pechal_superconducting_2018} and  atomic ensembles~\cite{treutlein_bose-einstein_2007}.  

The first step in the transfer process is an optical-to-mechanical state transfer, with a subsequent transfer to the final computing node~\cite{hease_bidirectional_2020,shandilya_optomechanical_2021,brubaker_optomechanical_2022}. An optical cavity is employed to enhance the radiation pressure during the optical-to-mechanical state transfer. Leading proposals work only in the {\it resolved sideband regime}, where  the decay rate of this cavity is lower than the mechanical resonance frequency~\cite{zhang_quantum-state_2003,palomaki_coherent_2013}. By contrast, most optomechanical systems operate in the {\it unresolved sideband regime}~\cite{aspelmeyer_cavity_2014}. In many cases this is due to the benefits that low mechanical frequencies convey for applications, for instance in precision sensing \cite{basiri-esfahani_precision_2019,sansa_optomechanical_2020, abbott_observation_2009}. In others, it is because of the difficulty of simultaneously achieving a low decay rate, a high resonance frequency, and sufficient radiation pressure coupling~\cite{leijssen_nonlinear_2017}.

To date, the only proposals for optical-to-mechanical state transfer in the unresolved sideband regime have used pulsed, rather than continuous, optomechanical interactions~\cite{vanner_pulsed_2011,bennett_quantum_2016,hoff_measurement-induced_2016}. This narrows the range of applications, introduces significant technical challenges due to the additional timing and phase accuracy required~\cite{vanner_cooling-by-measurement_2013,hoff_measurement-induced_2016, khosla_quantum_2017}, and involves large radiation pressure impulse forces that can be problematic~\cite{bennett_quantum_2016,muhonen_state_2019,bennett_quantum_2020}.

It is well known that a mechanical resonator can be feedback cooled close to its motional ground state in the unresolved sideband regime~\cite{mancini_optomechanical_1998}.  
Here we propose a continuous optical-to-mechanical state transfer protocol based on the same concept. By modelling the open quantum system dynamics, we show that feedback cooling can be understood as the transfer of a vacuum state of light onto the mechanical resonator. We find that appropriate choice of the feedback parameters allows the  transfer of arbitrary quantum states. 
The requirements for successful transfer closely match those for ground-state cooling -- once the optomechanical cooperativity exceeds the thermal occupancy of the mechanical resonator, a coherent state can be transferred with near unity fidelity and  the Wigner-negativity of non-Gaussian states can be preserved. Moreover, the feedback parameters can be used to phase-sensitively amplify (or {\it squeeze})  the transferred state, to engineer its temporal profile, and -- in direct analogy to state-transfer via resolved sideband cooling~\cite{peterson_laser_2016} -- to achieve the transfer of  a single optical sideband.

Our work extends continuous optomechanical state transfer beyond the resolved sideband limit, to low-quality optical cavities and low frequency mechanical resonators. Feedback cooling of a mechanical resonator to near its motional ground state has recently been demonstrated, both in cryogenic  \cite{rossi_measurement-based_2018} and room temperature environments~\cite{guo_feedback_2019}. As such, our proposal can be directly implemented with existing technology, providing a new tool for quantum networks and opening a new pathway to create and study macroscopic quantum systems. Our work also provides new insights into feedback cooling, showing that the process is in fact a quantum state transfer from light to mechanical motion.

\begin{figure}
    \centering
    \includegraphics[width=0.48\textwidth]{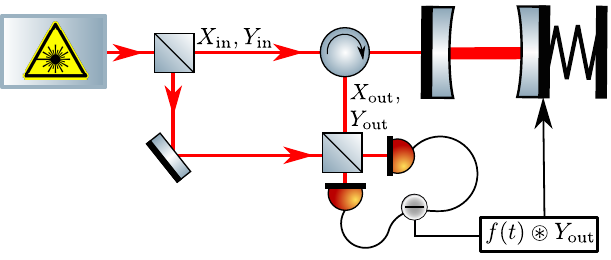}
    \caption{Schematic optomechanical system with feedback. Light is coupled into an optomechanical cavity. The reflected light is measured through homodyne detection. The detected photocurrent ($Y_{\rm{out}}(t)$) is convolved with a filter $f(t)$ and directly fed back to the momentum of the mechanical resonator.}
    \label{fig:schematic}
\end{figure}

We consider an optomechanical system in the unresolved sideband, high mechanical quality regime $(\kappa \gg \Omega \gg \Gamma )$ with resonant optical driving, where $\kappa \; (\Gamma)$ is the optical (mechanical) energy decay rate, and $\Omega$ the mechanical resonance frequency. In this scenario, the amplitude quadrature of the input optical field $X_{\rm in}$ is directly imprinted on the mechanical motion via radiation pressure. The phase quadrature $Y_{\rm in}$ is not, but is encoded on the phase quadrature of the output optical field as
\cite{bowen_quantum_2015}:
\begin{equation}
    Y_{\rm out} = - \sqrt{\eta} Y_{\rm in} + 2 \sqrt{\eta \Gamma C} Q + \sqrt{1-\eta} Y_{\rm v},
\end{equation}
where $\eta$ is the detection efficiency,  $C = 4g_{\mathrm{om}}^2 / \Gamma \kappa$ is the optomechanical cooperativity with $g_{\mathrm{om}}$ being the coherent-amplitude-boosted optomechanical coupling rate, $Y_{\rm v}$ is the vacuum noise introduced due to detection loss,
$Q$ ($P$) is the dimensionless mechanical position (momentum) operator with $[Q,P]=i$, and all optical quadrature operators are normalised such that $[X (t), Y(t')] = i \delta (t - t')$. We propose to detect the output phase quadrature and
use continuous feedback to transfer it to the mechanical resonator, as shown in Fig.~\ref{fig:schematic}. 
We note that feed-forward, similar to our feedback, has been applied to improve microwave-to-optical state transfer in the resolved sideband regime~\cite{higginbotham_harnessing_2018}. 
In contrast,  the feed-forward functioned in that experiment to suppress correlated noise terms, while both optical  quadratures were transferred by radiation pressure.  

Our scheme is analogous to feedback cooling~\cite{mancini_optomechanical_1998,cohadon_cooling_1999,hopkins_feedback_2003,poggio_feedback_2007,lee_cooling_2010,harris_feedback-enhanced_2012,rossi_measurement-based_2018,guo_feedback_2019}, with the detected signal applied as a force onto the mechanical resonator. Using quantum Langevin equations, we find that it is described by the following equations of motion:
\begin{equation}\label{Eqn:Qdot}
    \dot{Q} = \Omega P - \frac{\Gamma}{2}Q + \sqrt{\Gamma}Q_{\mathrm{in}},
\end{equation}
and 
\begin{align}
    \label{Eqn:Pdot}
    \dot{P} =& - \Omega Q - \frac{\Gamma}{2}P + \sqrt{\Gamma}P_{\mathrm{in}} - 2 \sqrt{\Gamma C} X_{\mathrm{in}} \\
    &- \frac{\Gamma G}{2} f(t)\circledast \left( -\left(Y_{\mathrm{in}} - \sqrt{\frac{1-\eta}{\eta}}Y_{\mathrm{v}}\right)\frac{1}{2\sqrt{\Gamma C}} + Q\right), \nonumber
\end{align}
where $P_{\rm in}$ and $Q_{\rm in}$ are white thermal noise operators that satisfy $[
Q_{\rm in} (t), P_{\rm in}(t')] = i \delta (t - t')$, and we have made the rotating wave approximation (RWA) with respect to the mechanical bath \cite{doherty_quantum_2012,bowen_quantum_2015}.
The last term of Eq.~\eqref{Eqn:Pdot} represents the feedback force, where the measured photocurrent is convolved with an arbitrary causal filter function $f(t) \in \mathbb{R}$ and amplified by the gain factor $G$. The filter function is normalised so that $|f(\Omega)|=1$, where $f(\omega) = \int_{-\infty}^{\infty} f \td e^{\I \omega t} \mathrm{d}t $ is the Fourier transform of $f(t)$.

The steady-state solutions to Eqs~\eqref{Eqn:Qdot}~and~\eqref{Eqn:Pdot} are found by moving into frequency space and adiabatically eliminating the dynamics of the optical cavity field (Supplementary Material, Section I \cite{Supp}).
This results in the quadratures
\begin{widetext}{
\begin{align}
\label{Qw}
    Q\fd &=  \sqrt{\Gamma}\chi(\omega) \bigg[ Q_{\mathrm{in}} +  \phi(\omega)P_{\mathrm{in}}
    - 2\sqrt{C}\phi(\omega) X_{\mathrm{in}} + \frac{G f(\omega)\phi(\omega)}{4\sqrt{C}} \left( Y_{\mathrm{in}} - \sqrt{\frac{1-\eta}{\eta}}Y_{\mathrm{v}}
    \right) \bigg], \\
\label{Pw}
        P\fd &= \sqrt{\Gamma}\chi(\omega)\bigg[ P_{\mathrm{in}} 
     - \left(\frac{G f(\omega)\Gamma}{2\Omega} + 1 \right)\phi(\omega)Q_{\mathrm{in}}
     - 2\sqrt{C} X_{\mathrm{in}} + \frac{G f(\omega)}{4\sqrt{C}} \left ( {Y_{\mathrm{in}}} - \sqrt{\frac{1-\eta}{\eta}}Y_{\mathrm{v}}
    \right ) \bigg],
\end{align}}
\end{widetext}
where
\begin{equation}
    \phi(\omega) = \frac{\Omega}{\Gamma/2 - i \omega},
\end{equation}
the feedback-broadened mechanical susceptibility is
\begin{equation}
    \chi(\omega) = \frac{1}{\Omega\phi(\omega)^ {-1} + (\Omega + G \Gamma \frac{f(\omega)}{2})\phi(\omega)}, \label{chi}
\end{equation}
and the adiabatic elimination is valid in the unresolved sideband regime ($\left\{\Omega, \;C \Gamma\right\} \ll \kappa$) taken throughout this paper.
From Eq.~(\ref{chi}), we see that the mechanical susceptibility decreases as $G$ increases. This suppresses most of the mechanical terms in Eqs~(\ref{Qw})~and~(\ref{Pw}). The only term that remains is $Q_{\rm in}$ in $P(\omega)$, but this is suppressed by the large mechanical quality factor $(\Omega/\Gamma \gg 1)$. It is this combined suppression of all mechanical terms that enables optical state transfer with high fidelity.

The optical input field consists of a continuum of optical modes. To build insight into which of these modes is best transferred to the single mechanical mode, as well as the gain and noise of the transfer process, we re-write Eqs~(\ref{Qw})~and~(\ref{Pw}) as:
\begin{eqnarray}
Q &=& g_X X_{\rm trans} + Q_{\rm noise,optical} + Q_{\rm noise,mechanical} \label{Qdef}\\
P &=& g_Y Y_{\rm trans} + P_{\rm noise,optical}+ P_{\rm  noise,mechanical}. \label{Pdef}
\end{eqnarray}
Here, $X_{\rm trans}$ and $Y_{\rm trans}$ are the optical quadratures  transferred to position and momentum, respectively, and $g_X$ and $g_Y$ are the transfer gains. Terms labelled with a subscript `noise' encompass the residual thermal variance remaining after feedback, and any optical terms not arising from the temporal mode of interest (\textit{i.e.}, inefficient detection, mode mismatch).

The input optical quadratures transferred to $Q$ and $P$ in Eqs.~(\ref{Qw})~and~(\ref{Pw}) are not perfectly conjugate observables. The difference is embodied in $\phi$, and is a result of the retarded response of the mechanical position to an applied force. The imperfection introduces an ambiguity in the optical mode that is optimally transferred -- a different mode is best transferred to $P$ and $Q$.  Here, we choose to assess the transfer of the mode that is optimally transferred to $P$. This mode is described by the annihilation operator
\begin{equation}
    a_{\rm trans}(\omega) = u(\omega) a_{\rm in}(\omega) \label{a_eqn}
\end{equation}
and spectral modeshape
\begin{equation}
    u(\omega) = \frac{2  \sqrt{\Gamma C}}{g_Y} \chi(\omega) \left (\frac{G f(\omega)}{8 C} - i \right), \label{modeshape}
\end{equation}
where $a_{\rm in} (\omega) = (X_{\rm in} (\omega) + i Y_{\rm in} (\omega))/\sqrt{2}$.
Using the relations $X_{\rm trans} =(a_{\rm trans}^\dagger + a_{\rm trans})/\sqrt{2}$ and $Y_{\rm trans}=i (a_{\rm trans}^\dagger - a_{\rm trans})/\sqrt{2}$, its amplitude and phase quadratures are found to be
\begin{eqnarray}\label{Qsig}
    X_{\rm trans} &=& \frac{2\sqrt{\Gamma C}}{g_{Y}} \chi(\omega) \left( \frac{G f(\omega)}{8C} X_{\rm in} + Y_{\rm in}\right)\\
    \label{Psig}
    Y_{\rm trans} &=& \frac{2\sqrt{\Gamma C}}{g_{Y}} \chi(\omega) \left(-X_{\rm in} + \frac{G f(\omega)}{8C} Y_{\rm in}\right).
\end{eqnarray}
Comparison of Eq.~(\ref{Psig}) with Eq.~(\ref{Pw})
confirms that $Y_{\rm trans}$ is reproduced exactly in $P(\omega)$, scaled by the momentum gain  $g_Y$.

The phase quadrature transfer gain, $g_{Y}$, can be determined by enforcing the boson commutation relation $[a_{\rm trans}(t), a_{\rm trans}^\dagger(t)]=1$ on
Eq.~\eqref{a_eqn}; while that for the amplitude quadrature, $g_{X}$,  can be found by requiring that the optical noise on position commutes with both  $X_{\rm trans}$ and $Y_{\rm trans}$, \textit{i.e.}, $[Q_{\rm noise, optical}(t),X_{\rm trans}(t)] = [Q_{\rm noise, optical}(t),Y_{\rm trans}(t)] = 0$, where $Q_{\rm noise, optical}$ is obtained by rearranging Eq.~\eqref{Qdef}. Together, these give
\begin{align}
    g_Y &=  \left[ \frac{4\Gamma C}{2\pi}\int_{-\infty}^{\infty} \left| \chi (\omega )\right|^2 \left( \left| f(\omega )\right| ^2+1 \right)  \mathrm{d}\omega \right] ^{1/2}\\
    g_X &= - \frac{1}{g_{\rm Y}}\frac{8\Gamma C }{2\pi} \int^{\infty}_{-\infty}|\chi(\omega)|^2 \Im{(\phi(\omega))\Im{(f(\omega))}} \mathrm{d}\omega.
\end{align}

The spectral modeshape and  quadratures of the transferred mode depend on both the feedback-broadened mechanical susceptibility $\chi(\omega)$ and the feedback filter function $f(\omega)$, so that the transferred state can be controlled through appropriate choice of the filter properties. Thus far our results are valid for an arbitrary real-valued causal filter function. In the remainder of the paper we choose the generalized-Lorentzian filter
\begin{equation}\label{Eqn:filter}
    f(\omega) = \frac{\Gamma' \Omega}{\omega^2 - \Omega^2 + \I \Gamma^{\prime} \omega},
\end{equation}
where $\Gamma^{\prime}$ is the filter bandwidth.  This filter is commonly used for feedback cooling~\cite{mancini_optomechanical_1998,pinard_full_2000,poggio_feedback_2007,harris_feedback-enhanced_2012} and is close to the known optimal filter for both momentum estimation~\cite{meng_mechanical_2020} and feedback cooling~\cite{doherty_feedback_1999}. $\Gamma^{\prime}$ is chosen to be much larger than $\Omega$, so that the filter  acts as an integrator near the mechanical resonance frequency. The gain factor $G$ can then be understood as the fractional increase in the mechanical decay rate due to the feedback.

With the filter in Eq.~(\ref{Eqn:filter}) and in the limit of large filter bandwidth and mechanical quality factor ($\Omega/\Gamma \gg 1$), the amplitude and phase quadrature transfer gains can be approximated as
\begin{equation}\label{Eqn:gs}
    g_{Y} = 2 \sqrt{C \left( \frac{ 1 + \frac{G^2}{64C^2}} {2+G} \right) } \, \, \, \, \text{and} \, \, \, \, g_{X} = \frac{1}{g_{Y}(1 + 2/G)}.
\end{equation}

\begin{figure}
    \centering
    \includegraphics[width=0.48\textwidth]{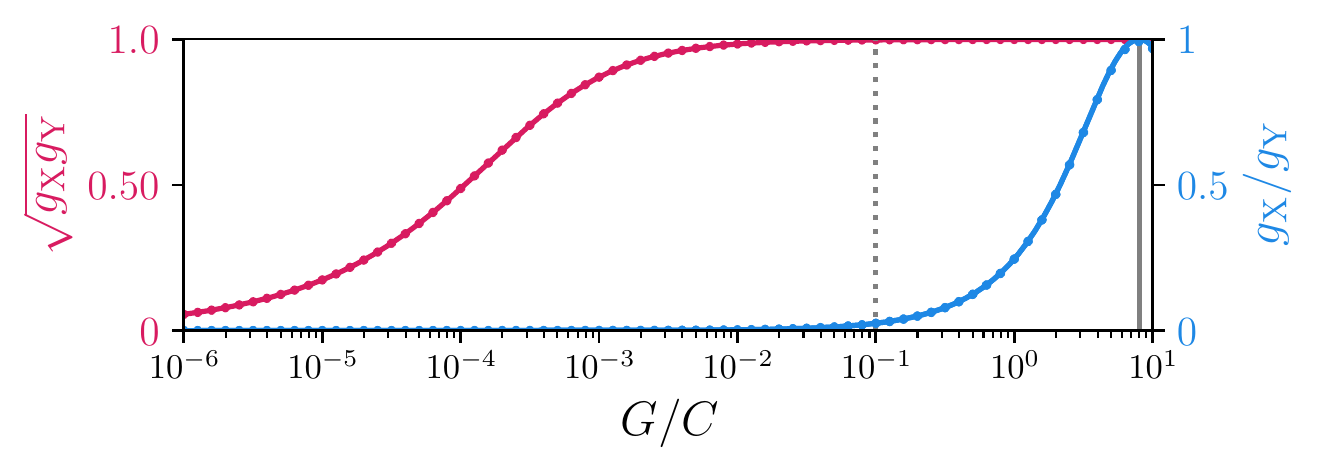}
    \caption{Transfer gain $(\sqrt{g_{X}g_{Y}},\,{\rm red})$ and squeezing $(g_{X}/ g_{Y},\,{\rm blue})$ as a function of the feedback strength by cooperativity $(G/C)$. The dashed line indicates G = 1 and the full grey line indicates the optimal gain $(G = 8C)$, where $g_{X}/g_{Y} = 1$. The dots are numerically obtained, and the lines are using the analytic expressions derived in the high quality factor limit.}
    \label{fig:gains}
\end{figure}

We define the overall gain of the transfer process as $\sqrt{g_X g_Y}$, so that it is independent of unitary squeezing operations on the transferred state~\cite{bowen_unity_2003}, and define the level of squeezing applied during the transfer as $g_X/g_Y$. The overall gain and squeezing level are plotted as a function of the feedback gain factor $G$ in Fig.~\ref{fig:gains} using both  numerical calculations and the analytic approximations of Eqs~(\ref{Eqn:gs}). For these plots and throughout the paper we use the system parameters $\Omega/2\pi\,=1\,\mathrm{MHz}$, $\Gamma/2\pi = 1\,\mathrm{Hz}$, $\Gamma'/2\pi = 1.59\,\mathrm{MHz}$,  $\kappa/2\pi\,=\,100\,\mathrm{MHz}$, and $g_{\rm om}/2\pi = 395\,\mathrm{kHz}$, $T = 30\,\mathrm{mK}$ which have been achieved in a range of optomechanical experiments~\cite{leijssen_nonlinear_2017,rossi_measurement-based_2018,guo_feedback_2019}.

The overall transfer gain approaches unity for $G \gg 1$, and the transfer generally involves amplitude quadrature squeezing ($g_X/g_Y <1$). Only at $G=8C$ do we find that the input state is transferred without any squeezing ($g_X/g_Y = 1$). 
Comparison of Eq.~(\ref{Qsig}) with Eq.~(\ref{Qw}) shows that, in the high quality limit for which $f(\omega)$ can be substituted with $f(\pm\Omega) = \mp i$, this choice of gain also results in near-agreement between $X_{\rm trans}$ and the optical input terms in $Q$. 
The remaining discrepancy arises from the retardation factor $\phi(\omega)$, and this discrepancy approaches zero in the high-quality-factor limit. We therefore select $G=8C$ for the remainder of the paper.

It is illustrative to consider how our choice of filter function and gain factor influences the spectral modeshape $u(\omega)$. The frequency dependence of the prefactor in Eq.~\eqref{modeshape} depends only on $\chi(\omega)$, and is sharply peaked at both $\pm \Omega$. However, since $f(\pm \Omega) = \mp i$, for $G=8C$ the term in parentheses is precisely zero at $- \Omega$ and equals $-2 i$ at $\Omega$. Our particular choice, therefore, enables a single-sideband state transfer, transferring only the lower optical sideband and doing this with a modeshape given approximately by $\chi(\omega)$ (see also Supplementary Material, Section II \cite{Supp}).

\begin{figure}
    \centering
    \includegraphics[width=0.48\textwidth]{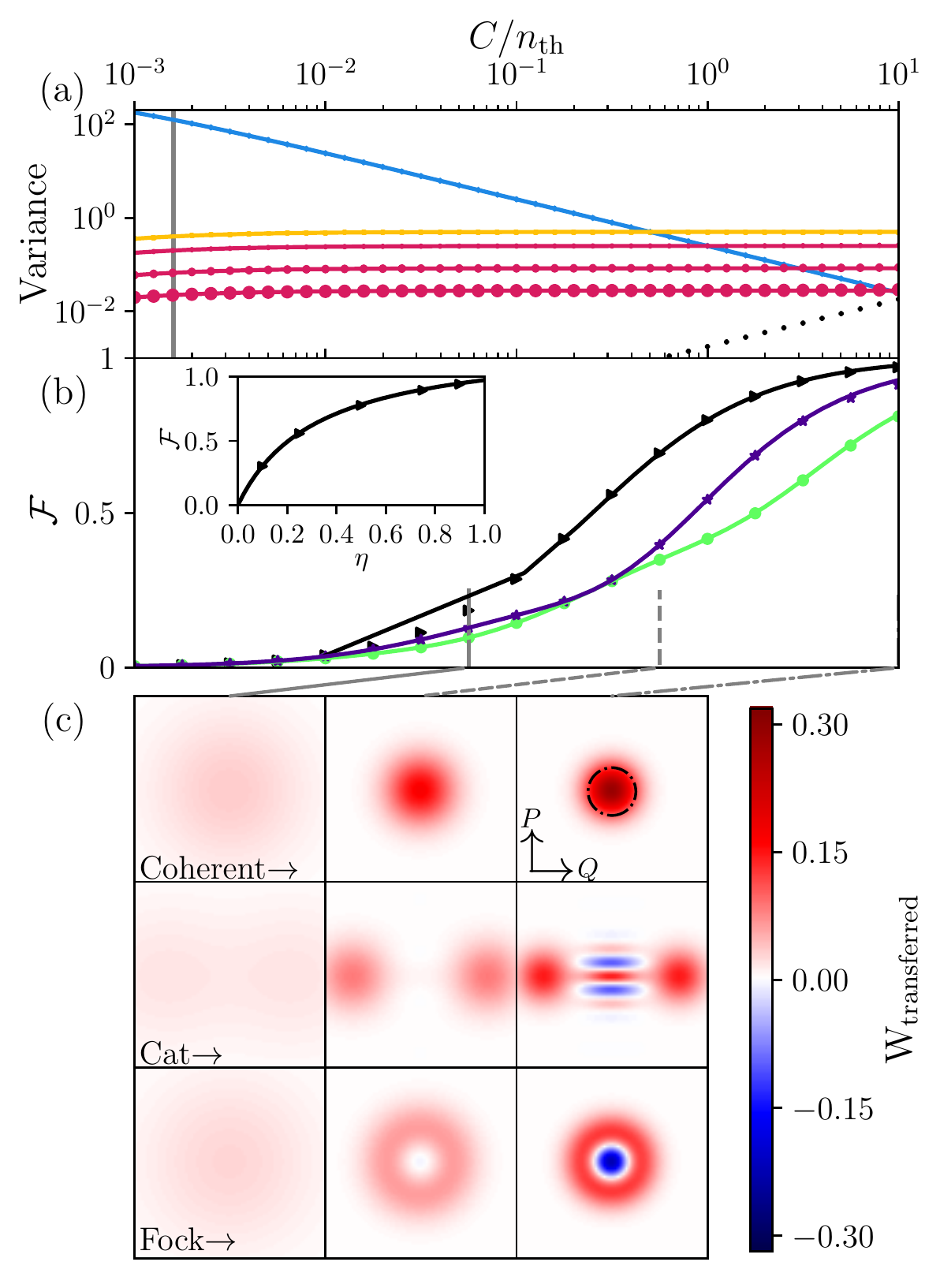}
    \caption{(a) Contributions to the variance as a function of interaction strength of mechanical noise (blue), optical signal (yellow), and two contributions of optical noise: mode mismatch on $Q$ (black), and inefficiency (red). The size of the markers correspond to the inefficiency ($\eta =\, 0.9,\, 0.75,\, 0.5$) for decreasing size respectively. (b) The transfer fidelity $( \mathcal{F} )$ as a function of interaction strength for a coherent state (black), cat state (green) $( \alpha = 2 )$ and single photon Fock state (dark blue). Inset shows $\mathcal{F}$ as a function of $\eta$ for the coherent state, at a fixed value of $C/n_{\mathrm{th}} = 10$. (c) Corresponding plots of the Wigner distributions for a coherent state (top row), cat state (middle row) and  Fock state (bottom row) at the interaction strengths indicated by the grey lines connected to subplot (b). The black dotted circle in the top right indicates the length scale of the contour of the ground state. The orientation of the plots is indicated by the black arrows in the top right plot.
    }
    \label{fig:VarianceFidelityDensities}
\end{figure}

To quantitatively assess the quality of transfer we first consider an input vacuum state. We calculate the contributions to the position and momentum variances 
from this input and from the noise sources specified in Eqs (\ref{Qdef})~and~(\ref{Pdef}) 
(see Supplementary Material, Sections II~\&~III \cite{Supp}). We separate the optical noise into contributions arising from inefficiences and  mode mismatch, so that the non-ideality of the transfer that arises due to $\phi(\omega)$ can be assessed.
The results are plotted in Fig.~\ref{fig:VarianceFidelityDensities}~(a) as a function of $C/n_{\rm th}$ (with $G = 8C$). The variance of the transferred optical mode increases with $C$, asymptoting to the vacuum variance of $1/2$ once $C \gg 1$.
Conversely, the mechanical noise contribution decreases, dropping below the vacuum level for $C \gg n_{\rm th}$. The variance of the optical inefficiency noise has a cooperativity dependence that is similar to the optical signal, increasing with $C$ and asymptoting to a constant value once $C \gg 1$. As expected, this noise increases as the detection efficiency degrades. However, even for $\eta$ as low as 0.5 the transferred signal variance still dominates inefficiency noise for the whole range of $C/n_{\rm th}$. The mode-mismatch noise on $Q$ is very low for small $C$, increases approximately linearly with $C$, and eventually exceeds the signal variance. Thus, the mode-mismatch ultimately constrains the performance of the state transfer. 

Using the analytic expressions for the gains in Eqs~(\ref{Eqn:gs}), we derive analytic expressions for the different variance contributions that are valid in the same high-quality, high-bandwidth limit (see Supplementary Material \cite{Supp}, Section III). With the exception of the mismatch noise, which is zero in the limit of high quality, these expressions agree well with the numerical results in Fig.~\ref{fig:VarianceFidelityDensities}~(a). From them, we find that when $C \gg 1$ the noise variance introduced by optical inefficiency is $V_\eta = (1-\eta)/4\eta$, and that the mechanical noise variance is suppressed below the vacuum noise level once $ C > \bar{n}_{\rm th}/2$.
 
Since the feedback process is linear and all noise sources are Gaussian, it is straightforward to extend our analysis beyond the transfer of vacuum states, to more elaborate states such as \Schrod{} cat states. This can be achieved using Wigner functions (Supplementary Material, Section IV \cite{Supp}). Imperfections introduced by the thermal noise, mode mismatch, and inefficiency tend to `smear out' quantum features of the transferred optical mode's Wigner function. Mathematically, this is represented by convolving the signal's Wigner function with a Gaussian noise kernel $\mathcal{G} (\textbf{r})$ (with $\textbf{r}= \left( Q \hspace{.2cm} P \right)^{T}$)~\cite{chountasis_quantum_1999}:
\begin{equation}
  W_{\mathrm{transferred}}  (\textbf{r})  =  \left( W \circledast \mathcal{G} \right) (\textbf{r}).
\end{equation}
In the regime relevant to this paper, $\mathcal{G}$ is typically close to symmetric, with a slight wider spread in the $Q$ direction due to mode mismatch.
The transfer fidelity can then be determined for any pure input state as
\begin{equation}
    \mathcal{F} = 2 \pi \int_{- \infty} ^{\infty} \int_{- \infty} ^{\infty} W(\textbf{r})  W_{\mathrm{transferred}}(\textbf{r}) \mathrm{d}^{2}\textbf{r} .
\end{equation} 

We plot the fidelity for input coherent, Fock, and cat states in Fig.~\ref{fig:VarianceFidelityDensities}~(b) as a function of $C/n_{\rm th}$ and assuming that $\eta=1$. The coherent state fidelity exceeds the classical limit of 1/2 at $C/n_{\rm th} = 0.25$ and the no-cloning bound of 2/3 at $C/n_{\rm th} = 0.50$. The fidelity for the non-Gaussian states also reach fidelities greater than 0.5 at similar, experimentally accessible~\cite{leijssen_nonlinear_2017,wilson_measurement-based_2015,rossi_measurement-based_2018} cooperativities. For the chosen experimental parameters, the maximum achievable  fidelities are 0.98, 0.93, and 0.82 for coherent, Fock, and cat states, respectively, and are limited by the mode-mismatch noise. The fidelity is robust against measurement inefficiencies as visible in the inset of Fig.~\ref{fig:VarianceFidelityDensities}~(b), which shows that the coherent state fidelity  can exceed 1/2 even with a detection efficiency as low as $\eta = 0.2$. Fig.~\ref{fig:VarianceFidelityDensities}~(c) plots the Wigner distributions of transferred coherent, Fock, and cat states at three different values of $C/n_{\rm th}$, showing that the negativity of the Fock and cat states can be transferred, and therefore non-classical properties of the input state preserved. 

In conclusion, we have identified that feedback can be used to achieve continuous optical-to-mechanical state transfer in the unresolved sideband regime. We predict that state transfer can be achieved with high fidelity and whilst preserving non-classical features such as Wigner negativity. The ability to implement continuous state transfer in the unresolved sideband regime significantly widens the class of optomechanical systems that can be used as interfaces in quantum networks.

\section*{Acknowledgements}
The authors thank Mr S. Khademi and Dr C. Meng for useful discussions. This research was primarily supported by the Australian Research Council Centre of Excellence for Engineered Quantum Systems (EQUS, CE170100009). Support was also provided by the by the Air Force Office of Scientific Research under award number FA9550-20-1-0391.

\section*{References}
\bibliographystyle{apsrev}

\clearpage
\pagebreak

\widetext
\begin{center}
	\textbf{\large Supplemental Material: Continuous optical-to-mechanical quantum state transfer in the unresolved sideband regime}
\end{center}
\setcounter{equation}{0}
\setcounter{figure}{0}
\setcounter{table}{0}
\setcounter{page}{1}
\makeatletter
\renewcommand{\theequation}{S\arabic{equation}}
\renewcommand{\thefigure}{S\arabic{figure}}
\renewcommand{\bibnumfmt}[1]{[S#1]}
\renewcommand{\citenumfont}[1]{S#1}

\section{Theoretical model}

The Quantum Langevin equation used to derive the equation of motion for an arbitrary quadrature ($O$) is \cite{bowen_quantum_2015}:
\begin{equation}
    \frac{\mathrm{d}O}{\mathrm{d}t} = \frac{1}{\I \hbar} \left[ O, H \right] - \left[O, a{^\dagger}\right] \left( \frac{\gamma}{2} a - \sqrt{\gamma} a_{in}(t)\right) +
    \left( \frac{\gamma}{2} a^{\dagger} - \sqrt{\gamma} a_{in}^{\dagger}(t)\right) \left[O, a\right],
\end{equation}
with $\gamma$ the linewidth and $a^{\dagger}$ ($a$) the creation (annihilation) operator associated with the operator $O$.

The full set of equations of motion (EOM) in frequency space, after including adiabatic elimination on the optical quadratures, is:
\begin{equation}
\begin{split}
    -\I \omega P &= -\Omega Q - \frac{\Gamma}{2} P + \sqrt{\Gamma} P_{\mathrm{in}} - 2\alpha g_{\mathrm{om,0}}X \\
    -\I \omega Q &= \Omega P - \frac{\Gamma}{2} Q + \sqrt{\Gamma} Q_{\mathrm{in}} \\
    0 &= -\frac{\kappa}{2} X + \sqrt{\kappa}X_{\mathrm{in}}\\
    0 &= -\frac{\kappa}{2} Y - \sqrt{\kappa}Y_{\mathrm{in}}-2\alpha g_{\mathrm{om},0} Q,
\end{split}
\end{equation}
where $\Gamma$ ($\kappa$) is the mechanical (optical) linewidth, and $g_{\mathrm{om,0}}$ the single-photon optomechanical interaction strength, which is boosted by $\alpha$ from the intracavity photon amplitude, $P_{\mathrm{(in)}}$ and $Q_{\mathrm{(in)}}$ are the  (intracavity) momentum and position quadratures of the mechanical oscillator, respectively, $\Omega$ is the mechanical resonance frequency and $X_{\mathrm{(in)}}$ and $Y_{\mathrm{(in)}}$ are the  (intracavity) amplitude and phase quadratures of the optical cavity respectively.

Using input-output relations $(O_{\mathrm{out}} = O_{\mathrm{in}} - \sqrt{\gamma}O)$, we can determine an estimate for the mechanical position quadrature, by rescaling the detected photocurrent:
\begin{equation}
    Y_{\rm out} = - \sqrt{\eta} Y_{\rm in} + 2 \sqrt{\eta \Gamma C} Q + \sqrt{1-\eta} Y_{\rm v},
\end{equation}
such that 

\begin{equation}
    Q_{\mathrm{est}} = Q - \frac{\sqrt{\kappa}}{4\alpha g_{\mathrm{om},0}} \left(Y_{\mathrm{in}} - \sqrt{\frac{1-\eta}{\eta}}Y_{\mathrm{v}} \right) ,
\end{equation}
 which includes the added noise $(Y_{\mathrm{v}})$ from detection inefficiencies $( \eta )$. After convolution with an arbitrary causal real-valued filter function $(f(t) \circledast Q_{\mathrm{est}})$ this is equivalent to an estimate of the momentum quadrature $(P_{\mathrm{est}})$. To model applying a feedback force we subtract $G \frac{\Gamma}{2} P_{\mathrm{est}}$ from the equation of motion for $P$, where $G$ is the strength of the feedback force. 
 
The steady state solutions for the mechanical quadratures are then as included in the main text (reproduced here for ease of reading):
\begin{equation}
\label{SQw}
    Q\fd =  \sqrt{\Gamma}\chi(\omega) \bigg[ Q_{\mathrm{in}} +  \phi(\omega)P_{\mathrm{in}}
    - 2\sqrt{C}\phi(\omega) X_{\mathrm{in}} + \frac{G f(\omega)\phi(\omega)}{4\sqrt{C}} \left( Y_{\mathrm{in}} - \sqrt{\frac{1-\eta}{\eta}}Y_{\mathrm{v}}
    \right) \bigg],
    \end{equation}
    and
    \begin{equation}
\label{SPw}
        P\fd  = \sqrt{\Gamma}\chi(\omega)\bigg[ P_{\mathrm{in}} 
     - \left(\frac{G f(\omega)\Gamma}{2\Omega} + 1 \right)\phi(\omega)Q_{\mathrm{in}}
     - 2\sqrt{C} X_{\mathrm{in}} + \frac{G f(\omega)}{4\sqrt{C}} \left ( {Y_{\mathrm{in}}} - \sqrt{\frac{1-\eta}{\eta}}Y_{\mathrm{v}}
    \right ) \bigg],
\end{equation}
with
\begin{equation}
    \phi(\omega) = \frac{\Omega}{\Gamma/2 - i \omega},
\end{equation}
and
\begin{equation}
    \chi(\omega) = \frac{1}{\Omega\phi(\omega)^ {-1} + (\Omega + G \Gamma \frac{f(\omega)}{2})\phi(\omega)}.
\end{equation}

Following the main text, we re-write Eqs.~(\ref{SQw})~and~(\ref{SPw}) in the form:
\begin{eqnarray}
Q &=& g_X X_{\rm trans} + Q_{\rm noise,~optical} + Q_{\rm noise,~mechanical} \label{SQdef}\\
P &=& g_Y Y_{\rm trans} + P_{\rm noise,~optical}+ P_{\rm  noise,~mechanical}. \label{SPdef}
\end{eqnarray}
Here, $X_{\rm trans}$ and $Y_{\rm trans}$ are the quadratures of the transferred optical mode that are imprinted on the position and momentum, respectively, and $g_X$ and $g_Y$ are the transfer gains allowing for possible differences in gain between position and momentum. Terms labelled with a subscript `noise' encompass the residual mechanical noise remaining after feedback, and the optical noise imprinted on the mechanical oscillator by both feedback and radiation pressure. 

Following the main text, we can derive the susceptibility of the transferred mode, by using the relations $X_{\rm trans} =(a_{\rm trans}^\dagger + a_{\rm trans})/\sqrt{2}$ and $Y_{\rm trans}=i (a_{\rm trans}^\dagger - a_{\rm trans})/\sqrt{2}$ to find:

\begin{equation}
    u(\omega) = \frac{2  \sqrt{\Gamma C}}{g_Y} \chi(\omega) \left (\frac{G f(\omega)}{8 C} - i \right) \label{Smodeshape}
\end{equation}

or in terms of $X_{\rm{trans}}$ and $Y_{\rm{trans}}$:

\begin{eqnarray}\label{SQsig}
    X_{\rm trans} &=& \frac{2\sqrt{\Gamma C}}{g_{\rm Y}} \chi(\omega) \left( \frac{G f(\omega)}{8C} X_{\rm in} + Y_{\rm in}\right)\\
    \label{SPsig}
    Y_{\rm trans} &=& - \frac{2\sqrt{\Gamma C}}{g_{\rm Y}} \chi(\omega) \left(X_{\rm in} - \frac{G f(\omega)}{8C} Y_{\rm in}\right).
\end{eqnarray}

The gain parameters are found through the method described in the main text:

\begin{eqnarray}
    g_Y &=&  \left( \frac{4\Gamma C}{2\pi}\int_{-\infty}^{\infty} \left| \chi (\omega )\right|^2 \left( \left| f(\omega )\right| ^2+1 \right)  \mathrm{d}\omega \right) ^{1/2}\\
    g_X &=& - \frac{1}{g_{\rm Y}}\frac{8\Gamma C }{2\pi} \int^{\infty}_{-\infty}|\chi(\omega)|^2 \Im{(\phi(\omega))\Im{(f(\omega))}} d\omega.
\end{eqnarray}

\section{Mechanical power spectral density}

We construct the symmetrised power spectral densities for all separate contributions listed in Eqs.~(\ref{SQdef})~and~(\ref{SPdef}) using the following relations:
\begin{equation}
\bar{S}_{\mathrm{OO}} = \frac{ S_{\mathrm{OO}}(\omega) +  S_{\mathrm{OO}}(- \omega)}{2},
\end{equation}
and
\begin{equation}
    \left<{Q_{\mathrm{in}}\Br{t}} Q_{\mathrm{in}}\Br{t^{\prime}} \right> = \left<{P_{\mathrm{in}}}\Br{t} P_{\mathrm{in}}\Br{t^{\prime}} \right>=\Br{\bar{n}_{\mathrm{th}} + 1/2}\delta\Br{t-t^{\prime}},
\end{equation}
where $\bar{n}_{\mathrm{th}} \approx k_B T/\hbar \Omega$ is the mean thermal occupancy of the mechanical resonator and $T$ is its temperature. Due to the typically high frequency of the optical field we approximate the optical field to have no thermal occupancy, with

\begin{equation}
    \left<{X_{\mathrm{in}}\Br{t}} X_{\mathrm{in}}\Br{t^{\prime}} \right> = \left<{Y_{\mathrm{in}}}\Br{t} Y_{\mathrm{in}}\Br{t^{\prime}} \right>=\delta\Br{t-t^{\prime}}/2.
\end{equation}

Using these, we find:
\begin{equation}
\begin{split}
S_{\mathrm{QQ,\, optical \, signal}} =&
     g_{Y}^2 \frac{1}{2 \pi }\frac{1}{2}4\Gamma \frac{4 g_{\mathrm{om}}^2}{\Gamma  \kappa }| \chi (\omega )| ^2  \left(| f(\omega )| ^2+1\right)  \\
S_{\mathrm{QQ,\, optical \, noise}}  =&
    \frac{1}{2}4\Gamma \frac{4 g_{\mathrm{om}}^2}{\Gamma  \kappa } 
    | \chi (\omega )| ^2 
    \Bigg [\left( \frac{1 - \eta}{\eta} \right) | f(\omega ) \phi (\omega )|^2 + \left| \phi(\omega)f(\omega) - g_{X} g_{Y}\right|^2 + \left|-\phi(\omega) - g_{Y}g_{X} f(\omega) \right|^2 \Bigg]
    \\
S_{\mathrm{QQ,\, mechanical \, noise}}  =&
\Gamma |\chi (\omega )| ^2 \left(1 + |\phi(\omega)|^2 \right)
    (\bar{n}_{\mathrm{th}}+\frac{1}{2})  \\
S_{\mathrm{PP,\, optical \, signal}} =&
    g_{Y}^2 \frac{1}{2}4\Gamma \frac{4 g_{\mathrm{om}}^2}{\Gamma  \kappa }| \chi (\omega )| ^2 \left(| f(\omega )| ^2+1\right)  \\
S_{\mathrm{PP,\, optical \, noise}}  =&
    \frac{1}{2}4\Gamma \frac{4 g_{\mathrm{om}}^2}{\Gamma  \kappa }| \chi (\omega )| ^2 \left(\frac{1-\eta}{\eta} \right)\left(| f(\omega )| ^2+1\right)  
    \\
S_{\mathrm{PP,\, mechanical \, noise}}  =&
    \Gamma | \chi (\omega )| ^2
    \left(1 + \left| 4\phi(\omega) f(\omega)  \frac{1}{\Omega} \frac{4 g_{\mathrm{om}}^2}{\kappa \Gamma} + 1 \right|^2 \right) \left( \bar{n}_{\mathrm{th}} + \frac{1}{2}\right) \\
S_{\mathrm{PQ}}  =&
    \frac{4 g_{\mathrm{om}}^2 }{\pi  \kappa }  | \chi (\omega )| ^2  \left(1 + \frac{| f(\omega )|^2 }{4} \left(1-\sqrt{\frac{1-\eta }{\eta }} \right)^2 \right) \Re{(\phi(\omega))} +\\
    & \frac{\Gamma}{2 \pi} |\chi(\omega)|^2
    \frac{8 g_{\mathrm{om}}^2}{\kappa\Omega^2 } \left(\Im\left(f(\omega \right)) \Im\left(\phi (\omega )\right)-\Re(f(\omega )) \Re(\phi (\omega ))\right)\left( \bar{n}_{\mathrm{th}} + \frac{1}{2}\right),
\end{split}
\end{equation}

\begin{figure}
    \centering
    \includegraphics[width=0.48\textwidth]{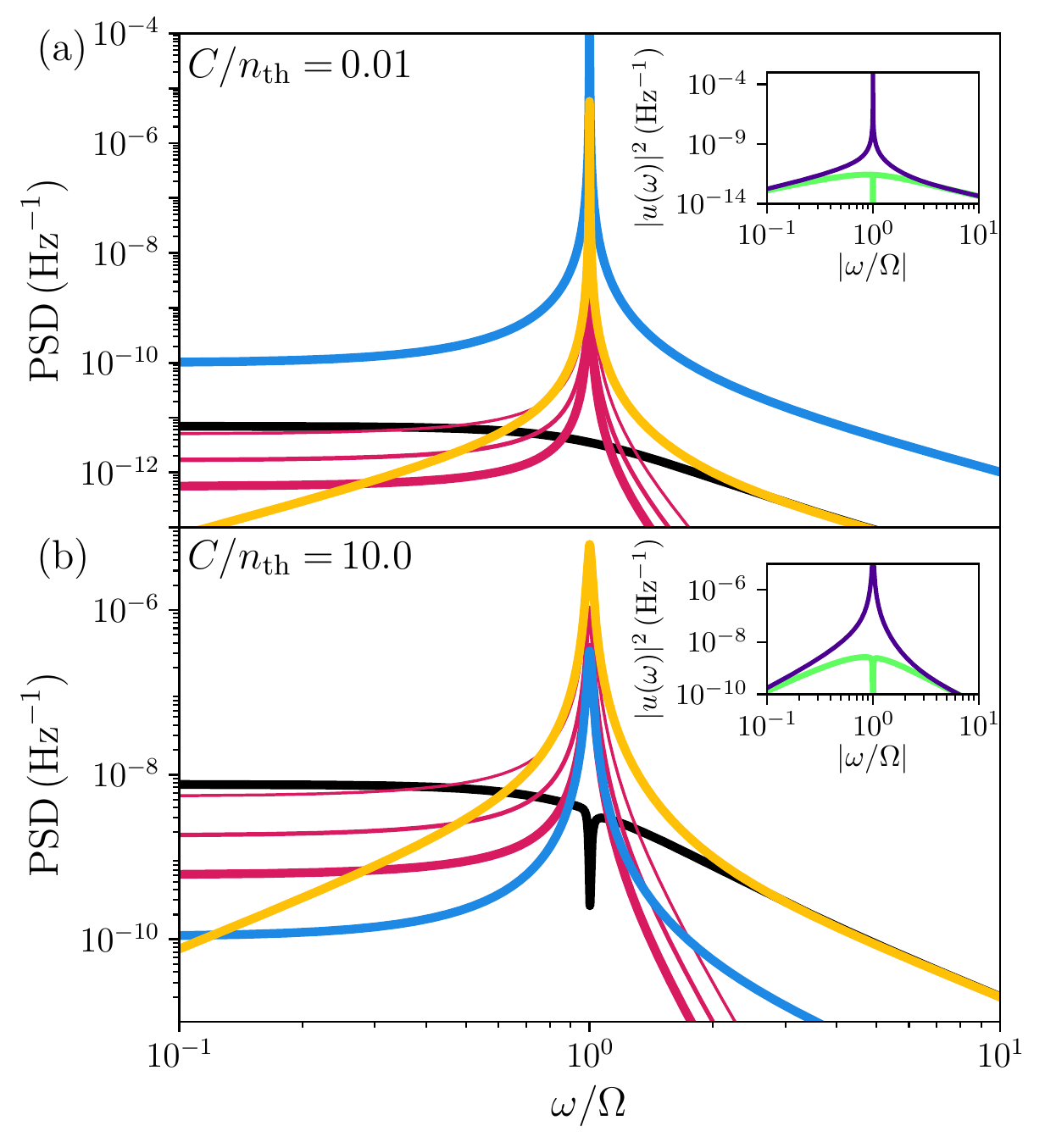}
    \caption{(a) Four separate contributions to the power spectral density (PSD) with $C/n_{\mathrm{th}} = 0.01$ of the mechanical quadratures: transferred optical mode (yellow), mechanical noise (blue), optical noise from mode mismatch on Q (black) and optical noise from inefficiencies $(\eta)$ (red). The width of the red lines correspond to $\eta =\, 0.9,\,0.75,\,{\rm and} \,0.5$ (decreasing width respectively). The inset shows the spectral mode of the signal for $\omega < 0 $ (green) and $\omega > 0$ (purple). (b) Same as in (a) but with a larger interaction strength, $C/n_{\mathrm{th}} = 10$.}
    \label{fig:PSDS}
\end{figure}

Each contribution is plotted for two different optomechanical cooperativities in Fig.~\ref{fig:PSDS}~(a)~and~(b), including inefficiency noise for three different detection efficiencies. Apart from the additional mode mismatch noise on $Q$, the contributions are near identical for $Q$ and $P$ in the limit that $C \ll \Omega/\Gamma$. We therefore only plot them for $Q$.

Apart from the mode-mismatch noise, all contributions to the power spectral density are peaked at the mechanical resonance frequency and have a roughly Lorenzian shape. The mismatch-noise, by contrast, is peaked at $\omega=0$ and approximates the shape of a low pass filter. At low cooperativity (Fig.~\ref{fig:PSDS}~(a)), the noise dominates the optical signal, preventing an effective quantum state transfer. At higher cooperativity (Fig.~\ref{fig:PSDS}~(b)),  the optical signal rises above the noise contributions, suggesting that quantum state transfer is possible. The inset in both figures plots the spectral modeshape of the  signal mode $(u(\omega))$. The optical signal peaks at $-\Omega$ (blue) and is suppressed relative to this peak by several orders of magnitude at $\Omega$ (green), evidencing that the chosen filter enables a single-sideband transfer.   

\section{Variances}

We obtain numerical values for the variances of each component of the power spectral density by integrating over frequency. To obtain the variance plot of the main text we sweep the interaction strength $g_{\mathrm{om}}$, while keeping the other system parameters constant. 

We also determine analytical approximations in the limit of a large bandwidth filter and high mechanical quality factor $({\rm Q} = \frac{\Omega}{\Gamma} \gg 1)$. In that limit, and with our chosen filter, 
the expressions for $g_{\rm Y}$ and $g_{\rm X}$ can be approximated as:
\begin{equation}\label{Eqn:SgPanalytic}
    g_{\rm Y} = \sqrt{2 C \left( \frac{ 1 + \frac{G^2}{64C^2}} {1+G/2} \right) },
\end{equation}
and
\begin{equation}\label{Eqn:SgQanalytic}
    g_{\rm X} = \frac{1}{g_{\rm Y}(1 + 2/G)}.
\end{equation}
Using these expressions and approximating the spectral densities in the high quality limit, we calculate analytic expressions for the variance of each component through the residue theorem. The results are: 
\begin{equation}\label{Eqn:SVQtrans}
  V_{{\rm X}_{\rm trans}} =  \frac{1}{2} {g_{\rm X}}^2,
\end{equation}
\begin{equation}\label{Eqn:SVQmech}
  V_{{\rm Q}_{\rm mech}} = \frac{1}{1 + G/2} \left( \frac{1}{2} + n_{\rm th} \right),
\end{equation}
\begin{equation}\label{Eqn:SVQmismatch}
  V_{{\rm Q}_{\rm mismatch}} = 0,
\end{equation}
\begin{equation}\label{Eqn:SVQeta}
  V_{\rm Q \, \eta} =  \frac{1}{4} \left( \frac{1-\eta}{\eta} \right) {g_{\rm X}}^2,
\end{equation}
\begin{equation}\label{Eqn:SVQmismatchsignal}
  V_{{\rm Q}_{\rm mismatch}{\rm Q}_{\rm signal}} = 0,
\end{equation}
and
\begin{equation}
  V_{\rm PQ} = V_{\rm QP} = 0.
\end{equation}
Eq.~\ref{Eqn:SVQmismatch} only occurs in the variance for $Q$. The analytic variances for $P$ are similar, with the subscript Q (X) substituted for P (Y).

\section{Wigner functions}

The Wigner functions used as optical input states are given by the following equations in Wigner space $( \mathbf{r} = (Q \hspace{.4pt} P)^T ) $:
\begin{equation}
    \begin{split}
        W_{\mathrm{Cat \,(Q= 2, P = 0)}}(\mathbf{r}) &= \frac{\exp (-\mathbf{r}\cdot\mathbf{r}) \left(\exp (-2 \alpha \cdot\alpha ) \cosh \left(2 \sqrt{2} \mathbf{r}\cdot\alpha \right)+\cos \left(2 \sqrt{2} \mathbf{r}\cdot(\varpi \alpha )\right)\right)}{\pi  (\exp (-2 \alpha \cdot\alpha )+1)} \\
        W_{\mathrm{Fock \, (n = 1)}}(\mathbf{r}) &= \frac{e^{-\mathbf{r} \cdot \mathbf{r}} \left(2 \mathbf{r} \cdot \mathbf{r}-1\right)}{ \pi } \\
        W_{\mathrm{Coherent,\,vacuum}}(\mathbf{r}) &= \frac{e^{-\frac{\mathbf{r}\cdot \mathbf{r}}{2} }}{2\pi},
    \end{split}
\end{equation}
where $\alpha = \left( \alpha_{\rm{r}}, \alpha_{\rm{i}}\right)$, and we use $\alpha_{\rm{r}} = 2$, and $\alpha_{\rm{i}} = 0$ for the calculation of the fidelity in the main text, and $\varpi = ((0,1),(-1,0))$ is a symplectic matrix.

As stated in the main text, the interactions are Gaussian and we can construct a Wigner function associated with the noise in the system:
\begin{equation}
    W_{\mathrm{noise}}(\mathbf{r}) = \frac{1}{2 \pi} \frac{1}{ \sqrt{ \det \left(V_{\mathrm{noise}}\right)} }
    \mathrm{exp}\left\{\frac{Q^2 V_{\mathrm{11}} - 2Q P V_{\mathrm{12}} + P^2 V_{\mathrm{22}}}{2V_{\mathrm{12}}^2 - 2 V_{\mathrm{11}} V_{\mathrm{22}} }\right\},
\end{equation}

where $V_{\mathrm{ij}}$ are the elements of the correlation matrix that only contains all the noise contributions. Specifically:
\begin{equation}
    V_{\mathrm{noise}} = 
    \left( \begin{array}{cc}
 V_{\mathrm{QQ,\,optical\,noise}} + V_{\mathrm{QQ,\,mechanical\,noise}} & V_{\mathrm{Q_{\rm{noise}}\,P_{\rm{noise}}}} \\
 V_{\mathrm{P_{\rm{noise}}\,Q_{\rm{noise}}}} & V_{\mathrm{PP,\,optical\,noise}} + V_{\mathrm{PP,\,mechanical\,noise}} 
\end{array} \right).
\end{equation}

The transferred Wigner functions $W_{\rm transferred}$ are found by:
\begin{equation}
   W_{\rm transferred} (\textbf{r}')  =  \left( W_{\mathrm{target}} \circledast W_{\mathrm{noise}} \right) (\textbf{r}),
\end{equation}
which we use directly for the calculation of the transfer fidelity:
\begin{equation}
    \mathcal{F} = 2 \pi \int_{- \infty} ^{\infty} W_{\mathrm{target}}(\textbf{r})  W_{\rm transferred}(\textbf{r}) d\textbf{r} .
\end{equation}

\section*{References}
\bibliographystyle{apsrev}

\bibliographystyle{apsrev}

\end{document}